\documentclass[conference]{IEEEtran}
\usepackage[hyphens]{url}
\usepackage{hyperref}
\hypersetup{breaklinks=true}
\hypersetup{draft}

\usepackage{comment}
\usepackage{enumitem}
\usepackage{multirow}
\usepackage{float}

\ifCLASSINFOpdf
\usepackage[pdftex]{graphicx}
\graphicspath{{../pics/}}
\DeclareGraphicsExtensions{.pdf,.jpeg,.png}
\else
\fi
\usepackage{tikz,ifthen}
\usepackage{verbatim}

\usepackage{amsmath,amsfonts,amsthm,bm} 
\newcommand\numberthis{\addtocounter{equation}{1}\tag{\theequation}}
\usepackage{algorithm}
\usepackage{algorithmic}

\usepackage[utf8]{inputenc}
\usepackage{fourier} 
\usepackage{array}
\usepackage{makecell}

\begin{document}
\bstctlcite{IEEEexample:BSTcontrol}
\raggedbottom
%
\title{MmWave Radar Point Cloud Segmentation\\using GMM in Multimodal Traffic Monitoring}
%
%
%

\author{
\IEEEauthorblockN{Feng Jin\IEEEauthorrefmark{1}, Arindam Sengupta\IEEEauthorrefmark{1}, Siyang Cao\IEEEauthorrefmark{1} and Yao-Jan Wu\IEEEauthorrefmark{2}}
\IEEEauthorblockA{\IEEEauthorrefmark{1}Department of Electrical and Computer Engineering\\
\IEEEauthorrefmark{2}Department of Civil and Architectural Engineering and Mechanics\\
The \textit{University of Arizona}, Tucson, AZ 85721\\
Email:\{fengjin, sengupta, caos, yaojan\}@email.arizona.edu}
} 
\maketitle

\begin{abstract}
In multimodal traffic monitoring, we gather traffic statistics for distinct transportation modes, such as pedestrians, cars and bicycles, in order to analyze and improve people's daily mobility in terms of safety and convenience. On account of its robustness to bad light and adverse weather conditions, and inherent speed measurement ability, the radar sensor is a suitable option for this application. However, the sparse radar data from conventional commercial radars make it extremely challenging for transportation mode classification. Thus, we propose to use a high-resolution millimeter-wave(mmWave) radar sensor to obtain a relatively richer radar point cloud representation for a traffic monitoring scenario. Based on a new feature vector, we use the multivariate Gaussian mixture model (GMM) to do the radar point cloud segmentation, i.e. `point-wise' classification, in an unsupervised learning environment. In our experiment, we collected radar point clouds for pedestrians and cars, which also contained the inevitable clutter from the surroundings. The experimental results using GMM on the new feature vector demonstrated a good segmentation performance in terms of the intersection-over-union (IoU) metrics. The detailed methodology and validation metrics are presented and discussed.
\end{abstract}

\begin{IEEEkeywords}
mmWave radar, radar point cloud, segmentation, Gaussian mixture model, classification, traffic monitoring.
\end{IEEEkeywords}

%
\IEEEpeerreviewmaketitle

\section{Introduction}
\par Using traditional radar signal processing, we obtain the position and Doppler information of reflection points from the scene after a suitable detection stage, such as Constant False Alarm Rate (CFAR) processing. The resulting positional representation in 3-D space is referred to as a radar point cloud, derived from a similar terminology used for LiDAR point cloud. Radar point cloud segmentation is a point-wise classification, which means it would classify each reflection point into a specific class. Segmentation for data obtained using camera (image or pixel array) and LiDAR (point cloud) have been continuously and extensively studied, primarily for autonomous driving and machine perception. Although relatively new, radar point cloud segmentation has also started to garner attention, given its several advantages over the other sensor modalities.

\par Traditional commercial radars offer limited resolutions, in both range and angle, which leads to a very sparse representation of the object from the radar's perspective. This also implies that segmentation on the sparse data is extremely difficult to model, often yielding sub-par results. On the other hand, camera and LiDAR provide a very dense pixel array and point cloud representation of the scene, respectively, that in turn yields a superior segmentation performance.

\par The recently emerging millimeter-wave (mmWave) frequency modulated continuous wave (FMCW) radar devices offer range resolution of up to 5 cm on account of an ultra-bandwidth of up to 4 GHz, using carrier frequencies of 60GHz, 77GHz and 80GHz, depending on the area of application. Furthermore, with advanced semiconductor fabrication process, more radio frequency (RF) channels are interpreted into a single monolithic microwave integrated circuit (MMIC) chip. This allows compact mmWave radars to provide relatively good angle resolution compared to outdated bulky commercial radars. Several examples of these mmWave FMCW MMIC radar chip include the Texas Instruments AWR1843 \cite{ref_AWR1843}, NXP TEF810X \cite{ref_NXP_radar1} and Infineon RXS816xPL \cite{ref_infineon_radar1}.

\par With the availability of such high-resolution radars, we can now obtain a relatively richer reflection point cloud representation of a single object, especially in the near range operation (less than 30 meters). Therefore, radar point cloud segmentation could be targeted by utilizing techniques from the traditional image and LiDAR processing domains. Furthermore, subsequent radar data post-processing, such as object clustering, tracking and classification, could be rebuilt using machine learning and deep learning architectures, similar to the ones used for images and/or LiDAR data, that have shown to yield very promising results.

\par Particularly, in multimodal traffic monitoring, sensors need to be employed to (i) estimate the traffic volume of different transportation modes, such as pedestrian, motorcycle and car, and (ii) estimate their average speeds. In order to achieve that, the sensor needs to be robust to operating all-day and in any weather condition with the additional capability to accurately estimate the speed of the objects, which makes radars a suitable choice. With the high-resolution relatively dense point cloud representation of each object, classification to a suitable transportation mode can be feasible by using a segmentation approach. 

\par In this paper, we use a single high-resolution mmWave radar device to monitor an experimental scene with pedestrian and car in it, and gather the radar point clouds. We propose to compute a new feature vector for each radar point. Then, we use a multivariate Gaussian mixture model (GMM) as the decision algorithm to perform the radar point cloud segmentation, i.e. point-wise classification. The structure of this paper is as follows: Section \ref{secII} presents a review of the current segmentation techniques; Section \ref{secIII} summarizes the multivariate Gaussian mixture model and the radar point feature vector we obtain from the mmWave radar point cloud data; Section \ref{secIV} presents the experimental setup and validation results; and Section \ref{secV} concludes this paper and discusses a future work.

\section{Literature Review of Segmentation Techniques}\label{secII}
\par
\begin{figure}[H]
\centering
\includegraphics[width=3.5in]{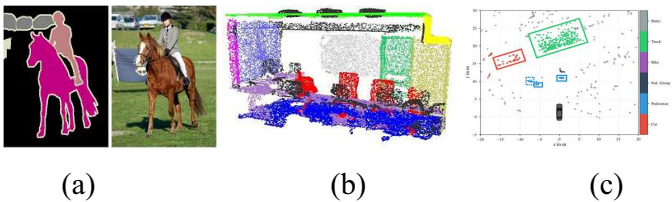}
\caption{Segmentation examples in which the color represents the class of object. (a) Image segmentation \cite{ref_imageseg_DL1}. (b) 3D (or LiDAR) point cloud segmentation \cite{ref_pointnet}. (c) Radar point cloud segmentation \cite{ref_radar_seg1}.}
\label{fig_segmentation}
\end{figure}

 This section reviews some latest segmentation techniques in the application domains of image, LiDAR and radar processing, as shown in Fig. \ref{fig_segmentation}.

\subsection{Image Segmentation}
\par Although image segmentation has had very broad approaches with a long researching history, those methods, such as thresholding-based, edge-based and region-based \cite{ref_imageseg_1}, heavily depend on the intensity (in grey or color) scale of each pixel. However, the radar cross-section (RCS), which is analogous to the intensity in the radar point cloud domain, may be too vague to be used. 

\par On the other hand, the clustering-based methods, such as $k$-means, Gaussian mixture model (GMM) and density-based spatial clustering of applications with noise (DBSCAN), that realize models to estimate the density or intensity-scale of the pixels, could be considered as valid choice for radar point cloud segmentation. Specifically, $k$-means assigns all the pixels into k clusters by minimizing the sum of the squared distance of all the pixels to its own cluster, as intuitively a cluster is thought of a group of data points whose inter-point distances are small compared to the distances to points outside of the cluster \cite{ref_patternbook}. The GMM models a group of data as a weighted sum of Gaussian distributions, where each distribution accounts for a unique cluster. A cluster, in this case, is formed if all the points obey the same Gaussian distribution \cite{ref_clusteringbook1}. In DBSCAN, a core point is defined if in its neighborhood of a given radius, there are at least a given minimum number of points. Then, the DBSCAN algorithm forms a cluster for all density-reachable points, i.e. each point is within the neighborhood of the core point, and all density-connected points, i.e. there is a third point from which both of these two points are reachable \cite{ref_clusteringbook1}.

\par Recently, deep learning based approaches have shown very promising results in image segmentation. In \cite{ref_imageseg_DL1}, the authors proposed a fully convolutional network (FCN) with end-to-end training on pixel-level labeled images. In \cite{ref_imageseg_DL2}, the authors proposed the R-CNN: regions with CNN features to first extract the ROI along with CNN features computation, and then to classify the region using a linear support vector machine (SVM). The success of supervised deep learning approaches motivates the researchers to apply it on the LiDAR point cloud segmentation.

\subsection{LiDAR Point Cloud Segmentation}
\par Each LiDAR point contains the information of 3D position and intensity. With a dense 3D point cloud representation of the object, the authors from Stanford proposed the PointNets family, including the PointNet \cite{ref_pointnet}, PointNet++ \cite{ref_pointnet++} and Frustum PointNet \cite{ref_frustum_pointnet}, to learn the 3D spatial feature of the object, which is a pioneering work on directly processing LiDAR point cloud, compared to the other traditional ways that may do the voxelization first and make the data unnecessarily voluminous. 
\par The authors first proposed a vanilla PointNet to transform the three-dimensional LiDAR point to the 1024-dimensional space in which the pattern of the different object can be more likely separable, according to the \textit{Cover's theorem} on the separability of patterns \cite{ref_haykin_neuralnetwork}. The basic architecture of the vanilla PointNet consists of multilayer perceptions to learn the feature space transformation in a supervised fashion with numerous labeled point data. And then, the authors devised the T-Net, a simplified vanilla PointNet, to learn the transformation of the object, such as translation, rotation and scaling, so that the entire PointNet architecture can be transformation invariant. 
\par In PointNet++, an extension of the PointNet, the authors introduced (i) the convolution operation with the PointNet as the kernel to learn the local spatial features, (ii) the multi-scale and multi-resolution grouping to deal with the variation in different areas, (iii) and the farthest point sampling (FPS) to sample the points in a more efficient way. 
\par Finally, in Frustum PointNet, the authors first used the typical convolution neural network (CNN) to detect the region of interest (ROI) in the 2D images, and then extracted the frustum of ROI in the 3D point cloud to represent the object following by a PointNet++ model to do the classification. The PointNet family can do one object classification and scene segmentation.

\subsection{Radar Point Cloud Segmentation}
\par Although segmentation in the synthetic aperture radar (SAR) image processing \cite{ref_seg_SAR} has been studied several years ago, segmentation on the radar point cloud has a very short history. This is on account of the previous real aperture radar's limited resolution, resulting in poor segmentation results, while the SAR has a relatively better resolution. 
\par With the great success of the PointNet family on LiDAR and 3D point cloud processing, researchers have attempted to try it out on radar point cloud, expecting promising results. In \cite{ref_radar_seg1}, the authors first accumulated multiple radar frames to obtain a richer point cloud, and then applied the Frustum PointNet with some minor adaptations on the 2D radar point cloud, and claimed better segmentation results over their previous work \cite{ref_radar_seg2} in which they used a combination of DBSCAN and long short-term memory (LSTM) network to predict the class for each radar point. And in \cite{ref_radar_seg3}, the authors applied PointNets on the 2D radar point cloud to differentiate the vehicle from clutter with the vehicle bounding box estimation. 
\par However, from our understanding, because the PointNet family is designed for learning the spatial 3D features of the object, it may not have a meaningful and practical results on the radar point cloud, as the radar point cloud is still very sparse compared to the LiDAR point cloud, it leads to the loss of some spatial features. Accumulation of multiple radar frames can improve the data. For a high-speed vehicle, however, its radar points would have moved a significant distance just after a few frames so that the accumulation does not make sense. Moreover, the availability of labeled radar point cloud is rare and difficult to gather, so the supervised learning approaches may not be a good option.

\section{Gaussian Mixture Model and MmWave Radar Point Cloud}\label{secIII}
Among all the available segmentation techniques from different application domains, we think the GMM model along with the relatively high-resolution mmWave radar data would be a feasible way to implement the radar point cloud segmentation.
\subsection{Basics of Multivariate Gaussian Mixture Model \cite{ref_patternbook}}
Given a set of data points in which each point is a vector, the goal is to classify each point into a single class. We assume there are a total of \(K\) classes these points may belong to. For a data point \(x\), given that it belongs to the \(k\)-th class, i.e. \(c_k=1\), \(k\in\{1,...,K\}\), it is assumed to follow a certain multivariate Gaussian distribution as 
\begin{equation}
p(x|c_k=1)=\mathcal{N}(\mu_k, \Sigma_k)=\frac{1}{(2\pi)^2|\Sigma_k|^{\frac{1}{2}}}e^{-\frac{1}{2}(x-\mu_k)^T{\Sigma_k}^{-1}(x-\mu_k)},
\label{gaussian}
\end{equation} 
where \(\mu_k\) is the mean and \(\Sigma_k\) is the covariance matrix for the \(k\)-th class. 
\par Then a data point with an unknown class should follow a GMM, which is a linear superposition of Gaussian distributions of all the \(K\) classes, by the \textit{law of total probability}, as
\begin{equation}
p(x)=\sum_{k=1}^{K}p(c_k=1)p(x|c_k=1)
\label{GMM}
\end{equation} 
where the \(p(c_k=1)\), also denoted as \(\pi_k\), is the prior probability of \(c_k=1\) or \(x\) belongs to the class \(c_k\).
\par If the parameters, i.e. \((\boldsymbol \pi, \boldsymbol\mu, \boldsymbol\Sigma)\) for all the \(K\) classes, are given, so the posterior probability of \(c_k=1\) for a given radar point, by the \textit{Bayes' theorem}, is 
\begin{align*}
\gamma(c_k)&={p(c_k=1|x)}\\
&=\frac{p(x|c_k=1)p(c_k=1)}{p(x)}=\frac{\pi_k\mathcal{N}(\mu_k, \Sigma_k)}{\sum_{j=1}^{K}\pi_j\mathcal{N}(\mu_j, \Sigma_j)},\numberthis
\label{posterior}
\end{align*}
where \(\gamma(c_k)\) can also be viewed as the responsibility that the class \(k\) takes for `explaining' the data point \(x\).
\par Then, we can use the maximum a posterior (MAP) criterion to determine the class of each radar point, that is
\begin{equation}
k=\max_{j}\;\gamma(c_j),\;\;\;j\in\{1,...K\},
\label{predict}
\end{equation}
\par Thus, the remaining question is how to determine all the parameters in GMM. The expectation-maximization (EM) algorithm can be applied as following. Assuming a set of data points \(X\{x_1, ..., x_N\}\) with unknown classes is collected, the optimal parameters \((\boldsymbol \pi^o, \boldsymbol \mu^o, \boldsymbol \Sigma^o)\) are those to maximize the likelihood function based on the maximum likelihood estimation (MLE), given by
\begin{equation}
ln\;p(X|\boldsymbol \pi, \boldsymbol \mu, \boldsymbol \Sigma)=\sum_{n=1}^{N}ln\;{p(x_n)}=\sum_{n=1}^{N}ln{\sum_{k=1}^{K}\pi_kp(x_n|c_k=1)},
\label{likelihood}
\end{equation}
\par The optimal parameters occur when the partial derivative of the likelihood function with respective to each parameter is zero. Then we have
\begin{align*}
\mu_k^o&=\frac{1}{N_k}\sum_{n=1}^{N}\gamma(c_k^n)x_n,\numberthis\\
\Sigma_k^o&=\frac{1}{N_k}\sum_{n=1}^{N}\gamma(c_k^n)(x_n-\mu_k^o)(x_n-\mu_k^o)^T,\numberthis\\
\pi_k^o&=\frac{N_k}{N},\numberthis
\label{max}
\end{align*}
where \(N_k=\sum_{n=1}^{N}\gamma(c_k^n)\), and \(c_k^n=1\) means the \(n\)-th point belongs to the \(k\)-th class.
\par
\begin{figure}[H]
\centering
\scalebox{0.5}{
\input{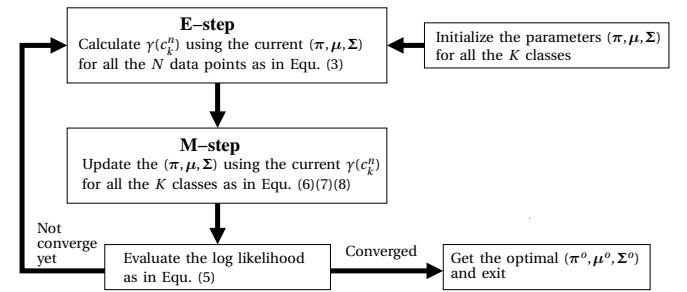}}
\vspace*{-0.1cm}
\caption{The EM algorithm.}
\label{fig_EM}
\end{figure}

 So the EM algorithm, as in Fig. \ref{fig_EM}, will iteratively update the parameters until the convergence of either the parameters or the log likelihood has been achieved.

\subsection{MmWave Radar Point Cloud in Multimodal Traffic Monitoring}
After the traditional FMCW radar signal processing chain, i.e. range-FFT, Doppler-FFT, angle-FFT, moving target indication (MTI), constant false alarm rate (CFAR), clustering and tracking, we obtain the radar point cloud, in which each point is a vector \(x(r, \theta_{az}, \theta_{el}, vD, snr, noise, pX, pY, pZ, vX, vY, vZ)\). Its parameters are listed in Table \ref{tab_raw_data}, in which the point data represents the radar measurement of each radar reflection point in the polar coordinate, and the centroid data represents the Kalman filtering based tracking results of the centroid of each tracked object in the Cartesian coordinate.
\vspace{-0.3cm}
\begin{table}[h]
\caption{Radar Point Cloud Data}
\label{tab_raw_data}
\centering
\scalebox{0.9}{
\begin{tabular}{|c|c|c||c|c|c|}
\hline
\multicolumn{3}{|c||}{\bfseries{Point Data}} & \multicolumn{3}{c|}{\bfseries{Centroid Data}} \\
\hline
\bfseries Symbol & \bfseries Value & \bfseries Unit & \bfseries Symbol & \bfseries Value & \bfseries Unit \\
\hline
\(r\) & range & m & \(pX\) & x position & m\\
\hline
\(\theta_{az}\) & azimuth angle & degree & \(pY\) & y position & m\\
\hline
\(\theta_{el}\) & elevation angle & degree & \(pZ\) & z position & m \\
\hline
\(vD\) & Doppler velocity & m/s & \(vX\) & x velocity & m/s \\
\hline
\(snr\) & Signal-to-noise ratio & dB & \(vY\) & y velocity & m/s \\
\hline
noise & CFAR window noise & dB & \(vZ\) & z velocity & m/s \\
\hline
\end{tabular}}
\end{table}

\par Then we propose the feature vector \((\Delta x, \Delta y, \Delta z, \Delta D, \sigma)\) for each radar point, where 
\begin{align*}
\Delta x&=r*\cos(\theta_{el})*\sin(\theta_{az})-pX,\numberthis\\
\Delta y&=r*\cos(\theta_{el})*\cos(\theta_{az})-pY,\numberthis\\
\Delta z&=r*\sin(\theta_{el})-pZ,\numberthis\\
\Delta D&=vD-(vX,vY,vZ)\cdot(the\_point\_direction)\\
&=vD - [\cos(\theta_{el})*\sin(\theta_{az})*vX+\\
&\cos(\theta_{el})*\cos(\theta_{az})*vY+\sin(\theta_{el})*vZ)],\numberthis\\
\sigma&=10\log_{10} r^4 +snr+noise.\numberthis\\
\label{feature_vector}
\end{align*}
\vspace*{-1cm}
\begin{figure}[H]
	\centering
	\includegraphics[width=3.5in]{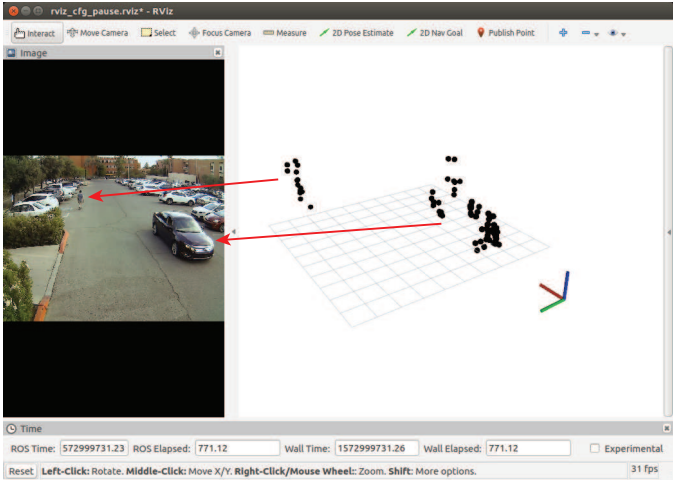}
	\caption{MmWave radar point cloud example. Red axis: x; Green axis: y; Blue axis: z.}
	\label{fig_pcl_example}
\end{figure}
\vspace{-0.3cm}
\par As result, (i) the \((\Delta x, \Delta y, \Delta z)\) is the relative position of each point with respect to the object centroid, and represents the extent of the object body, (ii) the \(\Delta D\) represents the relative Doppler, (iii) and the \(\sigma\) is the radar cross section (RCS) in the unit of dBsm. We observe that each point from one kind of object obeys a certain Gaussian distribution with its own mean and variance. In the multimodal traffic monitoring, because the size, speed and reflection coefficient of a pedestrian is distinguishable than these of a sedan, GMM can be applied for classification between these two. It is the same for the other transportation modes. Fig. \ref{fig_pcl_example} shows an example of radar point cloud including a car and a pedestrian from our data collection, that will be further described in Section \ref{secIV}. Here we can see the differences between the distributions of points from these two kind of objects.

\section{Experiment Setup and Field Test in Multimodal Traffic Monitoring}\label{secIV}
\subsection{Experiment Setup and Data Collection}\label{secExperiment}
\begin{figure}[H]
\centering
\includegraphics[width=3in]{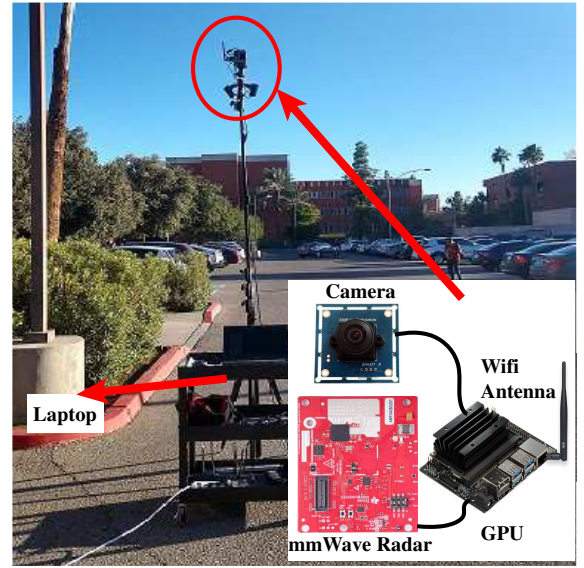}
\caption{Experimental setup.}
\label{fig_setup}
\end{figure}

We used a TI mmWave radar evaluation board AWR1843BOOST \cite{ref_AWR1843BOOST} to get the radar point cloud, the Nvidia Nano \cite{ref_NANO} to process the data, and one USB camera for capturing the video as a reference. Fig. \ref{fig_setup} shows how we collected the experimental data in a parking lot. The device was raised up to 3 meters high, and all the data was wirelessly transferred to a laptop for storage. 
\vspace{-0.3cm}
\vspace{-0.3cm}

\par With proper FMCW waveform design and the implementation of multiple-input-multiple-output (MIMO) direction-of-arrive (DOA) algorithm, we achieved about 9 centimeters of range resolution, 0.8 m/s of Doppler resolution, 15 degrees of azimuth angle resolution, and 28 degrees of elevation angle resolution. For now, we only collected the data with two different kinds of transportation modes, i.e. pedestrian and car. The effective radar detection area is up to 15 meters in range and 18 meters in cross-range, for both car and pedestrian. The data may also include the inevitable clutter or noise.
\par For the training data collection, because the GMM fitting is an unsupervised way, we kept one person continuously walking in the radar detection area, and one car driving through periodically. For the testing data collection, because the ground truth is needed to evaluate the model performance, so we let the person walking on the left side of the radar line-of-sight \((y=0)\), and the car driving on the right side. Then we labeled all the points with centroid \((y>0)\) as a pedestrian, all the points with centroid \((y<=0)\) as a car, and all the points without an associated centroid as clutter. 
\par Finally, we collected 8035 frames of training data with a duration of about 13 minutes, and 1222 frames of testing data with a duration of about 2 minutes.
\subsection{Experimental Results}
\par
\begin{figure}[H]
\centering
\includegraphics[width=3.5in]{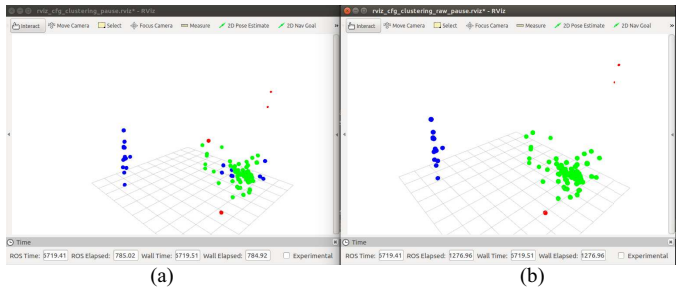}
\vspace*{-0.6cm}
\caption{One frame of results. Red point: clutter; Green point: car; Blue point: pedestrian. (a) Prediction. (b) Ground truth.}
\label{fig_results}
\end{figure}

 We used the scikit-learn APIs to fit the GMM using the training dataset, and saved the model to disk. Then we used the saved model to predict the testing dataset. Because the GMM fitting is an unsupervised approach, the GMM does not necessarily predict the same label as the ground truth. For example, the GMM may predict the pedestrian as an integer label, say, 0, while the ground truth for pedestrian would be other integer label, say, 1. So we visually associated the prediction label with the ground truth label. Fortunately, for a saved GMM model, this manual association just needs to be quickly done once. Finally, we evaluated the segmentation results. Fig. \ref{fig_results} shows one frame of the results. Referring to Fig. \ref{fig_pcl_example}, it is one example before the segmentation as all the radar points are colored in black, which means it has no class information.
\vspace{-0.3cm}
\vspace{-0.3cm}

\par To evaluate GMM in radar point cloud segmentation, we calculated the precision, recall and intersection-over-union (IoU) \cite{ref_PASCAL} as the performance metrics, as in the traditional image/LiDAR segmentation domain. In the interpretation of these metrics, the precision is intuitively the confidence that the model correctly classifies a point, and the recall is intuitively the confidence the model does not miss the detection of this object class. From the perspective of radar signal processing, high precision means a low false alarm rate; high recall means a low missed detection rate. Thus, a good model should have high precision and high recall simultaneously. And the F1 score, which is equal to \(\frac{2}{precision^{-1}+recall^{-1}}\), can be interpreted as one value metric of this model. The IoU, also called the Jaccard index, represents the percentage of overlap between the prediction and the ground truth. According to \cite{ref_PASCAL}, the IoU is recognized as the segmentation accuracy, and a model with $\geqslant$ 50\% overlap is considered good by standard. The results of GMM on the radar point cloud was presented in Table \ref{tab_iou}. As we can see here, the IoU of both pedestrian and car is above 50\%.
\vspace{-0.3cm}
\begin{table}[h]
\caption{Performance Metrics}
\label{tab_iou}
\centering
\begin{tabular}{|c|c|c|c|c|}
\hline
& \bfseries Precision & \bfseries Recall & \bfseries F1 Score & \bfseries IoU \\
\hline
Clutter & 0.71 & 0.89 & 0.79 & 0.66 \\
\hline
Car & 0.88 & 0.61 & 0.72 & 0.56 \\
\hline
Pedestrian & 0.85 & 0.93 & 0.89 & 0.80 \\
\hline
\end{tabular}
\end{table}

\par
\begin{figure}[H]
\centering
\includegraphics[width=2.5in]{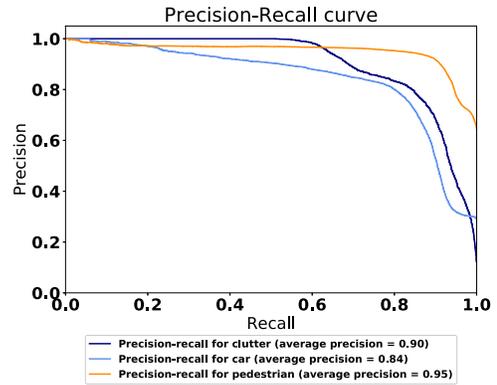}
\caption{Precision-recall curve.}
\label{fig_pr_curve}
\end{figure}

\begin{figure}[H]
\centering
\includegraphics[width=2.5in]{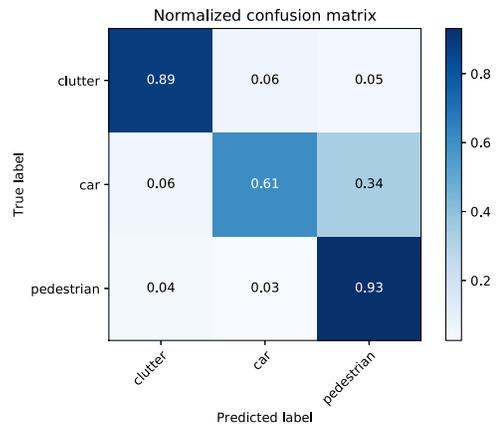}
\caption{Confusion matrix.}
\label{fig_cm}
\end{figure}

 To further evaluate the model performance, we plot the precision-recall curve as shown in Fig. \ref{fig_pr_curve}. In general, a point will be classified into class A, if the posterior probability of class A is greater than the $threshold=0.5$. As we adjust this $threshold$, the precision and recall changes accordingly. Normally, if we increase this probability $threshold$, the precision will be increased while the recall will be decreased; vice versa. The precision-recall curve shows the trade-off between these two. A good model has a position with both high precision and high recall. And we also computed the confusion matrix as shown in Fig. \ref{fig_cm}. 
\vspace{-0.3cm}
\vspace{-0.3cm}
\vspace{-0.3cm}
\vspace{-0.3cm}
\par Due to some difficulties in the data collection, the collected car data was less than the pedestrian data. Thus, the GMM model fitting was biased more on the pedestrian. And the performance of car point classification was relatively poor than that of the pedestrian. This can be alleviated if more data for both classes can be collected.
\par It is noted that this result is on a single frame basis, which means we do not accumulate multiple frames, such as in \cite{ref_radar_seg1}, as accumulation of multiple radar frames does not make sense in a scenario that vehicles are moving in high speed.

\section{Conclusion and Future Work}\label{secV}
\par In this study, we used a mmWave radar to capture the radar point cloud in which there are three kinds of objects, i.e. clutter, pedestrian and sedan. Then we implemented the GMM to perform the segmentation, i.e. the point-wise classification, and calculated the performance metrics such as precision, recall and IoU. And we found the GMM is simple but effectively achieves promising segmentation results. 
\par In the future, we aim to put the device at a traffic intersection to continuously collect more data with more transportation modes, such as pedestrian, motorcycle, bicycle, sedan, truck and bus, to further evaluate the GMM performance. As we expect, as the data complexity is increased, the simple GMM would fail to achieve a good performance. However, we will use the GMM as a preliminary classifier to help the DBSCAN algorithm, whose parameters are object-specific, to more robustly group the radar points from one object as one cluster. In return, the correctly clustered points will improve the object classification accuracy. So the work in this paper will be a part of our future work, which is to implement a joint clustering/tracking and classification in the multimodal traffic monitoring application using the mmWave radar sensor. 

\section*{Acknowledgement}
\par This project was funded by the National Institute for Transportation and Communities (NITC; grant number 1296), a U.S. DOT University Transportation Center, and also supported by Tucson Department of Transportation.

\ifCLASSOPTIONcaptionsoff
\newpage
\fi

\bibliographystyle{IEEEtran}
\bibliography{./bib/mydatabase}

\end{document}